\documentclass{eas}
\usepackage{graphicx}
\begin{document}

%%-----------------------------
%%      the top matter
%%-----------------------------
\title{planetary nebulae, 
tracers of stellar nucleosynthesis} 
\author{Gra\.zyna Stasi\'nska}\address{LUTH, Observatoire de Paris, CNRS, Universit\'e Paris Diderot ; Place Jules Janssen 92190 Meudon, France}

\begin{abstract}
We  review the information that planetary nebulae and their immediate progenitors, the  post-AGB objects, can provide to probe the nucleosynthesis and mixing in low and intermediate mass stars. We emphasize new approaches based on high signal-to-noise spectroscopy of planetary nebulae and of their central stars. We mention some of the problems still to overcome. We emphasize that, as found by several authors, planetary nebulae in low metallicity environments cannot be used to probe the oxygen abundance in the interstellar medium out of which their progenitors were formed, because of abundance modification during stellar evolution.
\end{abstract}
\maketitle
%%-----------------------------
%%      your text
%%-----------------------------
\section{Introduction}

Planetary nebulae (PNe) are produced from low- and intermediate-mass stars (LIMS) when, after having climbed the asymptotic giant branch (AGB) and blown away most of their envelope, they become very hot and able to ionize the expelled material. 

During the evolution of LIMS until the PN stage, recently manufactured nuclei are brought to the surface as a result of dredge-up processes, and modify the chemical composition of the star's outer layers that are being expelled. The underlying physical processes are complex, rich, and not fully understood (see e.g. Charbonnel 2005 or the contribution by Siess in this volume).
It is therefore natural to look for quantitative clues to these processes in  the chemical composition of planetary nebulae and of their central stars, which should trace the latest nucleosynthesis events in LIMS.

On the other hand, some elements are not expected to have their abundance modified in the envelopes of LIMS during the course of their evolution, and their abundances in PNe should then reflect the chemical composition of the matter out of which the progenitors were formed. Since the progenitors of PNe span a lookback time of over 1 Gyr, and possibly much more, PNe can be used as test-particles  to trace back the chemical evolution of galaxies. In this respect, they play the same role as luminous stars, with the advantage that the abundances of the elements are much easier to derive. They also require less telescope time to obtain spectra suitable for abundance analysis. Of course, elements whose abundance remain unchanged during the evolution of LIMS have first to be unequivocally identified. 

Unfortunately, PNe are still far from having fulfilled those expectations, and this for a variety of reasons. During the last decades, a number of PN studies have adressed both aspects mentioned above, with mitigated results. It is not only the conflicts in the derived abundances, but also the proper constitution of the samples and the large number of parameters influencing the changes in chemical composition that make such studies difficult. However, these investigations must be pursued and refined, in order to provide observational constraints to theoretical models of stellar evolution and nucleosynthesis as well as of chemical evolution of galaxies.

Recently, a few papers have adressed these topics with a somewhat general perspective (Perinotto et al. 2004, Gustafsson \& Wahlin 2006, Leisy \& Dennefeld 2006, Richer 2006, Stanghellini 2007), but they are by no means exhaustive. This review is obviously not complete as well. We wish to emphasize the diagnostic power of PNe as testbeds of LIMS nucleosynthesis, without discarding the involved difficulties. In Section 2, we present results from a few conventional analyses. In Section 3, we turn to discuss less conventional approches, some of which are just starting. In Section 4, we mention some of the problems that cannot be ignored. Conclusions are given in Section 5.

\section{Classical approaches to test nucleosynthesis of LIMS using planetary nebulae}

To start this section, it is fair to  say that, at large, planetary nebulae show obvious signatures of nucleosynthesis and mixing in their progenitors,  since,  at a given galactocentric distance, PNe exhibit a much larger dispersion  in their He/H and N/O ratios than HII regions (Henry et al 2000), and, on average, much higher values of those ratios.

\subsection{How to know which PN progenitors have experienced 2nd dredge-up?} 

 From theoretical models of stellar evolution,  the second dredge-up process, which brings H-burning products to the surface of the AGB star,  takes place only for stars with initial masses larger than about 4~M$\odot$, the exact value of this limit depending on the initial composition of the star (Iben \& Truran 1978, Becker \& Iben 1979). Peimbert (1978) proposed a classification of planetary nebulae in which  Type I  were defined as He- and N-rich PNe. He suggested that the excess of He and N was probably due to two causes:  the more massive nature of their progenitors, implying that they were formed more recently from a more chemically processes ISM, and a larger contamination of the ejected envelope due to stellar evolution. Since then, as reported by Torres-Peimbert \& Peimbert (1997), evidence has grown that Type I PNe correspond to the high mass end of PN progenitors. However, if Type I PNe are to be understood as those PNe whose progenitors have experienced second dredge-up, the definition of Type I should take into account the initial chemical composition of the progenitor. 
While the classical definition of Type I PNe is He/H $>$ 0.125 and  N/O $>$ 0.5 (Torres-Peimbert \& Peimbert 1997), these authors propose the  definitions He/H $>$ 0.105 and  N/O $>$ 0.38  for the LMC and He/H $>$ 0.10 and  N/O $>$ 0.22  for the SMC. Because of abundance gradients in the galaxy, the ISM value of N/O goes from 0.5 to 0.1 (Mart{\'{\i}}n-Hern{\'a}ndez et al. 2002), so, in principle, the definition of Type I PNe in the Milky Way should be dependent on Galactocentric radius. A similar concern should prevail when considering PNe in other galaxies. Some of this discussion can be found in Stasi\'nska et al. (1998) and Jacoby \& Ciardullo (1999) concerning PNe in the galaxies M31 and M32.

\subsection{Is the oxygen and neon abundance observed in PNe affected by nucleosynthesis in the progenitor star?}

Planetary nebulae have been used to study the chemical evolution of galaxies in which no other direct measurement of the oxygen abundance was possible (Richer et al. 1999). Obviously, this makes sense only if the oxygen abundance in planetary nebulae is identical to that of the interstellar medium out of which the progenitor star was born. Models of LIMS nucleosynthesis (Forestini \& Charbonnel 1997, Marigo 2000) show that the oxygen abundance in the astmospheres of these stars, and consequently in the PN envelopes, can be affected in both ways: reduction, due to the ON cycle, which converts oxygen into nitrogen, or enhancement resulting from $\alpha$-capture on carbon produced by the combustion of helium, followed by a 3rd dredge-up episode (at the highest  masses of PN progenitors however, hot bottom burning may significantly reduce the effect of  $\alpha$-capture on the final oxygen abundance). The fact that Ne/H and O/H are observed to correlate extremely tightly among large samples of planetary nebulae in the Galaxy, in the Magellanic Clouds and other galaxies (Henry 1990, Stasi\'nska et al 1998, Richer \& Mc Call 2006) and with the same slope as in HII regions is taken as an indication that the progenitors of PNe have not significantly modified the O and Ne abundances. On the other hand, from a comprehensive study of a large sample (183 objects) of PNe in the Magellanic Clouds, Leisy \& Dennefeld (2006) find that, neither oxygen nor neon can be used to derive the initial composition of the progenitor star, at least at the ``metallicity'' of the Small Magellanic Cloud. They draw this conclusion from the fact that N/O is seen to anticorrelate with O/H (especially for PNe in the SMC), and that the oxygen abundance in PNe is  found to be lower or higher than in HII regions. As previous authors, they do find a strong correlation between Ne/H and O/H, but they infer from it that the neon production is linked to oxygen. While their comparison with HII regions  would need to be redone, after a complete chemical composition analysis of a suitable sample of HII regions, it is wise, in the meantime, to be cautious with the use of PN oxygen and neon abundances in chemical evolution studies of low metallicity environments. Recently, Pe\~na et al. (2007) have performed a comparative analysis of both PNe and HII regions abundances in the Small Magellanic Cloud and in the irregular galaxy NGC 3109, confirming that the PN oxygen abundance is increased with respect to the initial abundance at low metallicities.

\subsection{PNe to test the importance of LIMS as sources of carbon production in galaxies}

That PNe progenitors produce and expell carbon is evidenced by the much larger  C/O ratios found in some in PNe with respect to HII regions of same O/H, as seen in Fig. 3a of Henry et al. (2000). Note however, that in the same figure, several PNe are seen to have their C/O much lower than any HII region, which seems difficult to understand in terms of stellar evolution, unless most of the carbon is actually in the form of dust grains. By comparing the abundance patterns of their sample of 20 PNe with homogeneously derived carbon abundances with yield predictions  from synthetic stellar evolution models (van den Hoek \& Groenewegen 1997, Marigo et al. 1996), Henry et al. come to the conclusion that the rough agreement between observations and models justifies confidence in the predicted LIMS yields to within a factor 10, but probably not better than a factor of 2 or 3. They, however, make the point that it is imperative to continue testing models with larger samples of PNe with accurately determined values of the carbon abundance, which requires spectroscopic observations in the UV. 

\subsection{Nucleosynthesis in extremely metal-poor environments: PNe in the Galactic halo}

The stars in the Galactic halo constitute an old, metal-poor population, and Galactic halo PNe should therefore provides observational clues to LIMS nucleosynthesis at low metallicities. There are only a dozen of halo PNe known so far. Strikingly, their C/O, N/O, Ne/O, S/O, Ar/O ratios show enormous scatter as compared to PNe in the Galactic disk (Howard \& Henry 1997). In the Howard \& Henry sample, which counts 9 objects, all the halo PNe have a N/O ratio larger than the solar value, while for the other elements there are several objects that show ratios smaller than solar. According to Howard \& Henry, all the trends are reasonably consistent with nucleosynthesis ideas relevant to intermediate-mass stars. At the same time, they note that part of the observed trends could be due to incomplete mixing in the halo. Clearly, advances in theoretical models since the study of Howard \& Henry, as well as the discovery of a few other halo PNe would  justify a new confrontation of the abundance patterns in halo PNe with theory. An extreme case is that of the recently discovered halo PN, PN G 135.9+55.9 (Tovmassian et al. 2001), the most oxygen-poor PN known, for which recent estimates (Stasi\'nska et al. 2005) give Ne/O twice solar, N/O three times solar, and C/O ten times solar for an O/H ratio between 1/40 and 1/100 of solar. 

\section{Less conventional analysis}

\subsection{Fluorine}
Three different scenarios have been proposed for the production of fluorine: explosions of Type II supernovae, stellar winds from Wolf-Rayet stars and third dredge-up in AGB stars. PNe thus allow one to test the latter scenario. Very high signal-to-noise spectra are needed for that. Zhang \& Liu (2005) have determined the abundance of fluorine in a sample of 14 PNe. They find an average abundance of fluorine of twice the solar value, the highest value being 12 times solar. They also find that F/O increases with C/O. These findings suggest that indeed  thermally pulsating AGB stars play  a role in the production of fluorine. Very few data on fluorine abundance are available in other astronomical sites, so it is not clear presently whether LIMS are the main sources of fluorine production. In some of the PNe they studied, Zhang \& Liu (2005) actually find a subsolar value of F/O. This could be explained by a destruction of F in the He intershell, through the $^{19}$F($\alpha,n$)$^{22}$Ne reaction. 

\subsection{$s$-process elements}
It has been understood since at least Burbidge et al. (1957)  that $s$-process elements are produced in evolved LIMS. Yet, the first identification  of $s$-process elements in a PN had to await several decades. P\'equignot \& Baluteau (1994) detected lines of various $s$-process elements in a deep optical spectrum of the brightest PN, NGC 7027. They also noted that, in fact, the [Kr IV] line appears to be present in half of the published PN spectra. This has fostered interest in detecting  $s$-process elements in PNe and measuring their abundances by yet other techniques. Dinerstein and her collaborators (Dinerstein 2001, Sterling et al. 2007) have started near IR studies of PNe to detect  fine-structure  lines from post-Fe 
peak ions. They have also used FUSE spectra of the central stars of PNe to detect resonance line of $s$-process elements in the intervening nebular shell (Sterling el al. 2002, Sterling \& Dinerstein 2003). Sharpee et al. (2007) performed echelle spectroscopy on four bright PNe. From all these studies, it results that Kr and Xe show abundance enhancements of typically  factors 10 over solar in many cases. However, solar-like values are found for a number of PNe. It is interesting that most of the $s$-process elements that can be detected in PNe are different from those that can be detected in the atmospheres of evolved stars. Therefore, PN spectroscopy  provides valuable  information, complementary to stellar spectroscopy on the production site of $s$-process elements. In the very near future, starting with  the survey of 120 PNe for the study of the abundances of Se and Kr (Sterling et al. 2007 and another paper to come), we will have a much clearer picture of the production of those elements in LIMS.

\subsection{Isotopes}

Obviously, stronger tests of LIMS nucleosynthesis would be achieved by determining the abundances of isotopes of various elements. 
In PNe, the only isotopic abundances that can be determined are those of  $^3$He from radio observations and of $^{13}$C from UV observations of the CIII]$\lambda$1909 line or millimetric observations of CO. Those observations are difficult, and only a few positive cases of detection or upper limits exist. Yet, they lead to interesting results.

In standard stellar models, $^3$He is produced by low mass stars. However, the coupling of these standard $^3$He yields with chemical evolution models of the Galaxy produce $^3$He/H ratios much higher than observed in Galactic HII regions (see Fig. 3 of the review by Prantzos 2005). Charbonnel \& Do Nascimento (1998) suggest  that 96\% of low mass stars are affected by an extra mixing process proposed by Charbonnel (1995), which is induced by rotation and destroys the freshly produced  $^3$He. The same process would increase the $^{13}$C abundance. It is remarkable that NGC 3242, one of the two PNe known to have a large  $^3$He/H (Balser et al. 2006) indicating its possible descendence from  one of the few stars that did not experience extra mixing, actually shows a $^{12}$C/$^{13}$C  ratio compatible with this hypothesis (Palla et al. 2002). In most other PNe where the  $^{12}$C/$^{13}$C  is  measured, this ratio is found to be lower than 15 (Palla et al. 2000, Balser et al 2002), while standard stellar models predict a ratio between 20 and 30, thus reinforcing the idea of the extra mixing proposed by Charbonnel. 

Rubin et al. (2004) retrieved archival IUE high-resolution spectra of PNe to  attempt  the measurement of the $^{13}$C line.  From a sample of about 50 promising objects, they could measure this line only in one object,  NGC 2440, which is thought to have an unusually high mass central star, and whose progenitor was likely more massive than 4 M$\odot$. They obtain $^{12}$C/$^{13}$C=4.4. This is compatible with the scenario of hot-bottom burning predicted for AGB stars with masses above  4 M$\odot$, which drives the $^{12}$C/$^{13}$C towards the CN cycle equilibrium value of 3.3.

\subsection{PN central stars}

In contrast to abundances in PN envelopes, abundances in PN central stars are available for a limited number of objects. The observational requirements are larger, since spectra of sufficient wavelength resolution are needed. Also, photospheric abundance analysis is less straightforward than nebular abundance analysis. The importance of studying the chemical composition of PN central stars is at least twofold. It allows one to compare the abundances of the same elements in the star and in the surrounding nebula (e.g. He, C, N, O) (see e.g. Napiwotzki et al. 1994, Rauch et al 2002). Note that those do not necessarily have to be equal, since the abundances in the nebular envelopes integrate a longer mass-loss time interval  than the present-day photospheric abundances of the central stars. The other advantage is that they allow one to determine the abundances of elements that either are not seen in  nebular spectra or may be depleted on dust grains in the nebular envelopes. 

The recent paper by Werner \& Herwig (2006) reviews the chemical abundances of H-poor PN central stars, i.e. stars of spectral types [WC] and PG1159 and compare them to predicted abundances from state-of-the-art models for AGB evolution and late He-shell flash. They find good qualitative and quantitative agreement. They mention that the relatively high H abundances in some [WC] stars can be explained by a post-AGB He-shell flash rather than a late He-flash. This would also explain the evolutionary sequence  from late-[WC] to early-[WC] central stars that accounts for the majority of PNe with [WC] type nuclei (G\'orny \& Tylenda 2000). However, one difficulty with this scenario is that  late-[WC] subtypes show  systematically higher C/He abundance ratios than early-[WC] subtypes. Abundance determinations in [WC] stars are presently being reexamined, by means of the last generation of non-LTE models for expanding stellar atmospheres which account for line-blanketing for iron-group elements  and wind clumping (Hamman et al. 2005, Todt et al. 2006).

As regards H-rich central stars of PNe, which represent about 90\% of cases (G\'orny \& Stasi\'nska 1995), the abundance studies are paradoxically yet quite sparse but systematic studies based on far UV data and modern stellar atmospheres are on-going (Werner et al. 2005, 2007).

\subsection{Post-AGB stars}

Post-AGB stars, the immediate precursors of planetary nebulae, i.e. stars that have achieved their super-wind phase and have left the AGB, but are not yet hot enough to ionize their surrounding nebula,   provide information complementary to that of PNe in the understanding of LIMS nucleosynthesis. 
As argued by Stasi\'nska et al. (2006), optical post-AGB stars have several advantages with respect to planetary nebulae: 1) abundances of a large variety of elements can be obtained, including those of $s$-process elements; 2) the abundance of carbon, a crucial element,  is known with the same degree of precision as that of  O and N, while in PNe, the carbon abundance is  significantly less reliable and more difficult to obain; 3) the determination of stellar effective temperature and gravity allows one to estimate the mass of the post-AGB star by direct comparison with theoretical stellar evolutionary tracks. On the other hand, the derivation of  the chemical composition of the atmospheres of such stars requires a very careful analysis of their atmospheres with  up-to-date stellar atmosphere models.
Stasi\'nska et al. (2006) opened the way  to the statistical use of post-AGB stars for testing LIMS nucleosynthesis,  with a sample of 125 objects for which the chemical composition was taken from the literature. In this pilot study, they only considered the abundances of C, N, O, Ne, S, Zn. The main results  were that:
a)  the vast majority of objects which do not show evidence of N production from primary C   have  a low stellar mass ($M_{\star}$ $<$ 0.56 M$_{\odot}$);  b) there is no evidence that objects which did not experience 3rd dredge-up have a different stellar mass distribution than objects that did; c) there is clear evidence that 3rd dredge-up is more efficient at low metallicity. The sample of known post-AGB stars is likely to increase significantly in the near future thanks to the   ASTRO-F and follow-up observations, making these objects even more promising as testbeds for AGB nucleosynthesis. The ``Toru\'n catalogue of Galactic post-AGB and related objects''  (Szczerba et al. 2007) with his evolving, on-line version (http://www.ncac.torun.pl/postagb) should help in the systematization of such studies.

\subsection{Analysis of PN abundances via tailored models for the TP-AGB phase}

To-date, the only detailed comparison of PN elemental abundances with model predictions is by Marigo et al. (2003). This work considers the abundances of He, C, N, O, Ne, S and Ar for a sample of 10 PNe with spectra available not only in the optical, but also in the IR and in the UV ranges, which allows a more accurate determination of abundances through the observations of many more ionization stages than available just from the optical. These abundances have been analyzed by comparison with TP-AGB models, adjusting the stellar parameters so as to reproduce the observed abundances. The parameters include the initial stellar mass and metallicity, molecular opacities, dredge-up and hot-bottom burning efficiencies. It was found that the sample of PNe could be divided in two classes. The first one is composed of PNe with low He content (He/H $<$0.15) and solar-like oxygen abundances. The second one is composed of He-rich (He/H $>$0.15) and O-poor PNe. The main results of this scrutineous investigation are the following. 

For the first group, progenitors are stars with masses in the range 0.9--4.0 M$\odot$ and solar-like chemical composition. There is evidence for carbon enrichment in some of the PNe. The carbon abundances are well reproduced with TP-AGB models with dredge-up efficiencies 0.3--0.4 and adopting molecular opacities related to the chemical composition of the envelopes. The nitrogen abundances are consistent with the expectations from the first and second dredge-up. The helium abundances are consistent with enrichment due to first and possibly second and third dredge-up, without the need of contribution from hot-bottom burning.

For the group of (He-rich, O-poor) PNe, progenitors are stars of subsolar metallicity with masses of 4--5 M$\odot$, experiencing both the third dredge-up and hot-bottom burning. The N/H -- He/H correlation and the N/H--C/H anti-correlation are quantitatively explained by the occurence of a number of very efficient, carbon-poor dredge-up events. The neon abundances imply a significant production of $^{22}$Ne during thermal pulses, thus reducing the role of the $^{22}$N($\alpha,n$)$^{25}$Mg reaction as a neutron source to the $s$-process nucleosynthesis in these stars. 

Marigo et al. (2003) note the apparent inconsistency between the  masses of  4--5 M$\odot$ for the progenitors, and the subsolar metallicitie of the PNe of the latter group. Normally, for such high progenitor masses,  one should expect metallicities  close to those of the interstellar medium, since the progenitors must have formed recently. They suggest that perhaps another physical process has not been fully taken into account.

\section{Sources of difficulties in the analysis}

\subsection{Abundance analysis}

Compared to the chemical composition of stellar atmospheres, the chemical composition of PNe is straightforward to obtain. One determines the electron temperature using temperature-sensitive line ratios ([OIII]4363/5007 or [NII]5755/6584 being the most popular ones). Then one derives the ionic abundances of the elements  dividing the observed line fluxes by the line emissivities (which depend on the temperature). Finally, elemental abundances are computed from the ionic abundances with the help of ionization correction factors.  The latter are generally obtained from analytical fits of photoionization models. The most used ones to-date are those of Kingsburgh \& Barlow (1994). There are however some problems in abundance determinations, that are not always fully appreciated (see Stasi\'nska 2004). Indeed, error bars in published abundances of the same objects do not always overlap, even if the abundances have been obtained by  tailored photoionization modelling. One big issue is whether an appropriate value of the electron temperature is been chosen for each ion. If, as argued by Peimbert \& Peimbert (2006), temperature fluctuations occur in the nebular gas, then these fluctuations must be corrected for in order to obtain an unbiased abundance. However, the existence of temperature fluctuations to a level that will affect abundance determinations is still an open issue (see Stasi\'nska 2007). Another, perhaps related concern, is the fact that in many PNe, abundances derived from recombination lines (RLs) are larger than abundances derived by the common methods, which are based on collisionally excited lines (CELs). The discrepancies are by factors of several, the most extreme case being a factor of 70 (see Liu 2006). The current interpretation of this discrepancy is that recombination lines come essentially from cold clumps of H-deficient plasma, embedded in the nebular shell. In the most extreme cases of RL/CEL abundance discrepancies, the abundances derived by common methods are heavily biased. However, it seems that in the case of RL/CEL discrepancies of a factor of 2--3, the abundances derived from the CELs give a fair estimate of the average heavy element abundances in the nebula. The physical origin of those cold clumps is not yet clear. In some cases (like in the PNe Abell 30 and 78), the H-deficient material has likely been ejected by the star during a helium flash. However, in most cases the chemical composition of those clumps is incompatible with theories of nucleosynthesis, since the solar C/N/O/Ne ratio is preserved. One appealing possibility is that the clumps might be the debris of planetesimals pertaining to the progenitor star (Liu 2003).

  \begin{figure} [h]
\includegraphics[scale=0.6]{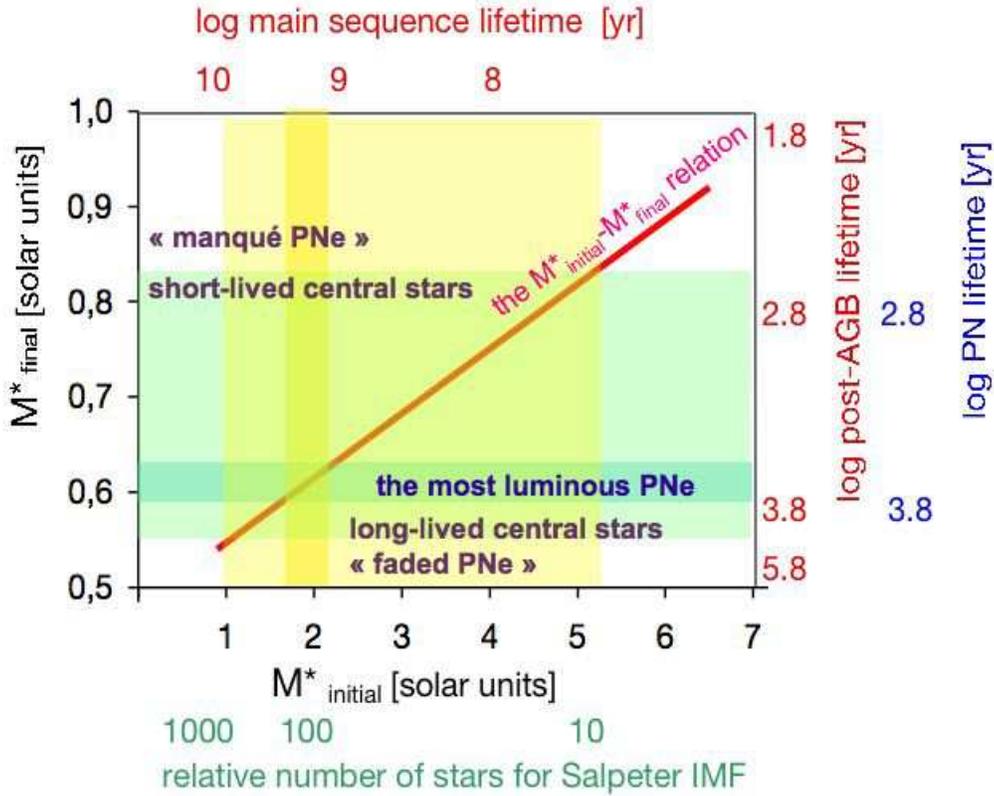}
    \caption{A multiparameter sketch of the classical view of the relation between PNe and their progenitors.}
    \label{sample-figure}
  \end{figure}

\subsection{Depletion on dust grains}
Planetary nebulae contain dust grains, and refractory elements can be partly or entirely in solid from. Therefore, abundances of such elements as Mg, Si, Fe, which can be determined in PNe,  are in fact useless to test LIMS nucleosynthesis. Even carbon can be importantly depleted. 

\subsection{The initial-final mass relation and sample biases}

Much of the interpretation of abundance patterns in PNe relies on a classical view of LIMS evolution, by which LIMS of given mass necessarily end up in a stellar core whose mass is entirely determined by the initial mass. This is the initial-final mass relation, considered to be universal (Weidemann 2000).

Figure 1 shows, in a schematic way, the implication of this initial-final mass relation on the population of PNe, taking into account that the post-AGB evolution is much faster for high core masses (the evolution time is roughly proportional to $M_*^{-9}$)\footnote {The numbers appearing in the figure are not to be taken litterally, as the figure was made ``by hand'', mostly for illustrative purposes}. The figure allows one to infer at what lookback time was the PN progenitor star formed, for a given present-day core mass. It also shows that PN studies miss both the highest initial masses (unless one is precisely observing the object during its extremely short life time as a PN) and the lowest initial masses (because the nebula would have dispersed before being ionized). 

However, the suggestion that  bipolar PNe would actually result only from LIMS that have evolved in close binary systems with common envelope phase (Livio 1993) has now been extended to the suggestion that \emph {most} PNe would derive from binaries (Moe \& de Marco 2006, Soker 2006). This would actually explain why we see so many PNe in the Galactic bulge while, according to the detailed study of Zoccali et al. (2003), all the stars in the Galactic bulge have formed at least 10 Myrs ago. Indeed, in the common-envelope scenario, the mass loss rate is higher and the evolution time of the ionizing star is shorter. If the binary evolution led to merging, this  could  perhaps make understandable the finding by Marigo et al. (2003) that their group of metal-poor PNe have high progenitor masses. 
On the other hand, if indeed PNe are produced by the evolution of binaries, then the interpretation of PN abundance patterns in terms of LIMS nucleosynthesis becomes much more difficult, since the present day masses of the central stars are not easily linked to the progenitors masses. 

\section{Conclusion}

We have briefly reviewed some of the information that PNe and their immediate progenitors, the so-called post-AGB objects, can provide to test our understanding of the nucleosynthesis of low and intermediate mass stars. While traditional approaches use the abundances of He, C, N, O and Ne measured in PNe envelopes, high signal-to-noise spectroscopy has recently allowed the determination of F and $s$-process elements abundances. Determination of chemical abundances in the atmospheres post-AGB stars and PNe central stars is developing, thanks to the use of large telescopes and modern stellar atmosphere codes. All this offers potential clues to nucleosynthesis and mixing processes occuring in LIMS. However, there are many difficulties on the way, not only because of uncertainties in abundance determinations, but also because the observational  samples are rather small while there are quite a few parameters determining the evolution of LIMSs (not only stellar masses and metallicities, but other factors such as binarity for example). By comparison of abundances in PNe and HII regions found in the same galactic environment, one can infer which of the elements observed in PNe can indicate the composition of the ISM where the parent stars were formed. It appears that oxygen is not a good tracer of the ISM oxygen abundance in low metallicity environments.
%%-----------------------------
%%      your bibliography
%%-----------------------------

\end{document}